\documentclass[prb,twocolumn,showpacs,preprintnumbers,amsmath,amssymb]{revtex4}
\usepackage{graphicx}
\usepackage{graphicx}
\usepackage{dcolumn}
\usepackage{bm}
\usepackage{braket}

\begin{document}

\title{Role of excited states in Shockley-Read-Hall recombination in wide-band-gap semiconductors}

\author{Audrius Alkauskas$^1$}
\email{audrius.alkauskas@ftmc.lt}
\author{Cyrus E. Dreyer$^2$}
\thanks{Current address: Department of Physics and Astronomy, Rutgers University, Piscataway, NJ 08845-0849, USA}
\author{John L. Lyons$^{2}$}
\thanks{Current address: Center for Functional Nanomaterials, Brookhaven National Laboratory, Upton, New York 11973, USA}
\author{Chris G. Van de Walle$^2$}

\affiliation{$^1$Center for Physical Sciences and Technology, Vilnius LT-01108, Lithuania\\
$^2$Materials Department, University of California, Santa Barbara, California 93106, USA\\}

\date{\today}

\begin{abstract}
Defect-assisted recombination is an important limitation on efficiency of optoelectronic devices. However, since nonradiative capture rates decrease exponentially with energy of the transition, the mechanisms by which such recombination can take place in wide-band-gap materials are unclear. Using electronic structure calculations we uncover the crucial role of electronic excited states in nonradiative recombination processes. The impact is elucidated with examples for the group-III nitrides, for which accumulating experimental evidence indicates that defect-assisted recombination limits efficiency. Our work provides new insights into the physics of nonradiative recombination, and the mechanisms are suggested to be ubiquitous in wide-band-gap semiconductors.

\end{abstract}


\maketitle

\begin{figure}
\includegraphics[width=8.5cm]{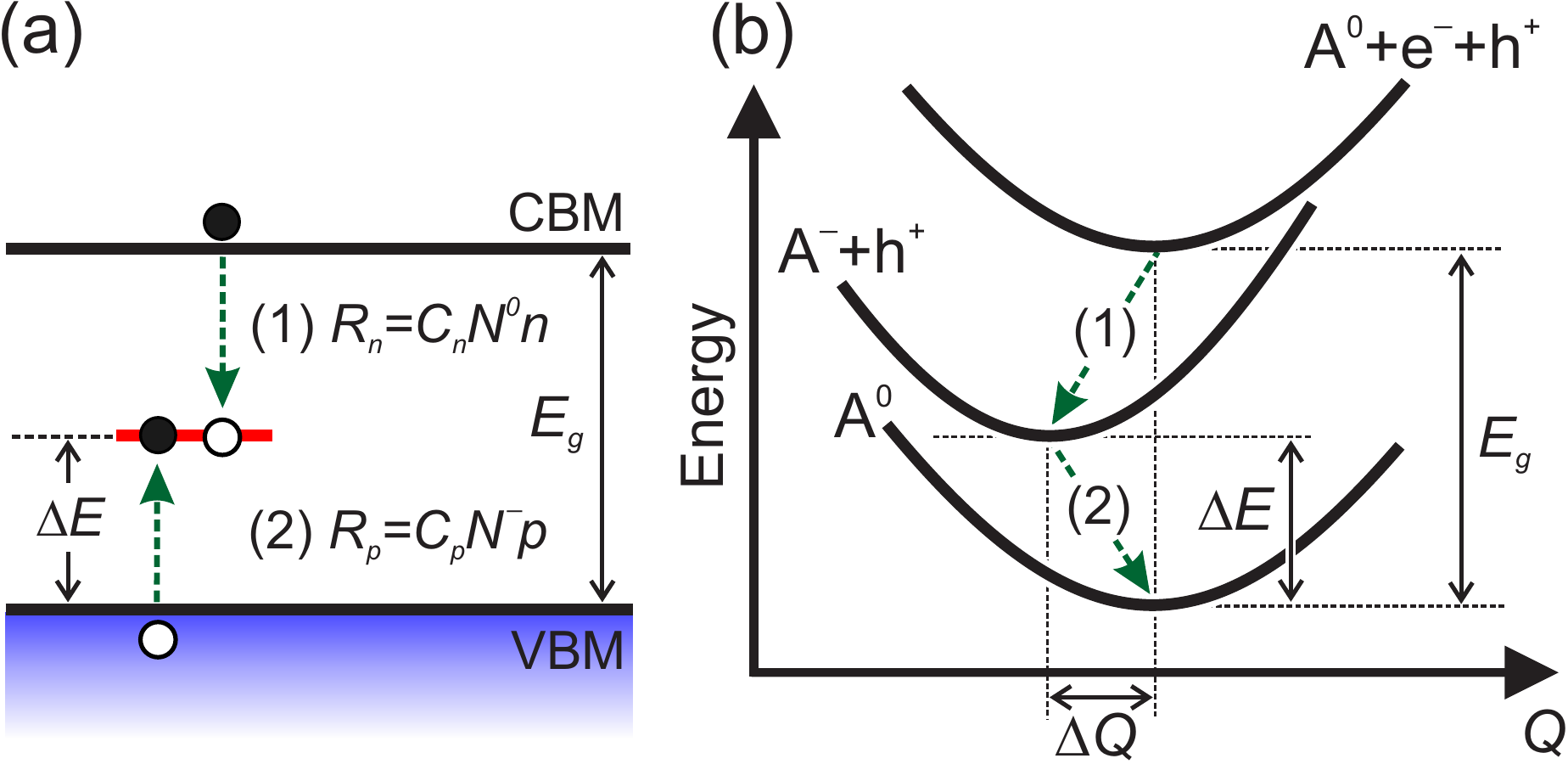}
\caption{(Color online) Shockley-Read-Hall recombination at a defect with a level at energy $\Delta E$ above the valence-band maximum (VBM), in two different representations: (a) energy level diagram; (b) configuration coordinate diagram.
Electron and hole capture rates $R_n$ and $R_p$ are expressed as a function of capture coefficients $C_n$ and $C_p$, carrier densities $n$ and $p$, and defect concentrations in
charge states $N^0$ and $N^{-}$. }
\label{SRH}
\end{figure}

Defect-assisted recombination is a process that limits the efficiency of many electronic and optoelectronic devices. The classic Shockley-Read-Hall (SRH) recombination picture considers a defect with a single energy level in the band gap \cite{Shockley_PR_1952,Hall_PR_1952}, as illustrated in Fig.~\ref{SRH} for a defect with charge states $0$ and $-1$. A complete recombination cycle requires capture of an electron by the neutral defect, followed by capture of a hole at the negatively charged defect. The overall recombination rate is thus governed by the {\it slower} of the two processes. Radiative capture rates are typically too small (an issue discussed in more detail below), and nonradiative capture rates decrease roughly exponentially with the energy of the transition \cite{Abakumov}; this trend would lead one to conclude that defect-assisted recombination becomes unimportant in wide-band-gap materials.

However, evidence is mounting for the occurrence of SRH recombination in materials with band gaps of 2.5 eV or larger \cite{David_APL_2010,Meneghini_APL_2014,Aleksiejunas_APL_2014}. Strong motivation for these studies has been provided by the technological importance of group-III nitride semiconductors as the key materials for light emitters in the green and blue spectral regions \cite{Nakamura_IEEE_2013}. With band gaps of 3.5 eV for GaN and 0.7 eV for InN \cite{Handbook}, InGaN alloys can cover the entire visible spectrum. Progress in nitride growth currently allows producing materials with dislocation densities $<$$10^6$ cm$^{-2}$, and there is a consensus that it is point defects that cause SRH recombination in such materials \cite{Aleksiejunas_APL_2014,Han_2015,Meneghini_2015}. However, the microscopic origin and mechanism of SRH recombination in nitrides has remained elusive. This is a serious handicap for improving device efficiencies.

In this work we show that intra-defect excited states can play a key role in enabling carrier capture required for SRH recombination. These excited states are derived from defect-related orbitals, as opposed to the shallow hydrogenic states that have been previously invoked to explain carrier capture at some charged centers in, e.g., Si or GaP \cite{Rees,Mott}. The wider the band gap of the material, the greater a role these excited states are likely to play. We demonstrate this for the highly relevant case of gallium vacancy complexes in nitride semiconductors, where processes involving excited states increase nonradiative recombination rates by many orders of magnitude. Such mechanisms play a key role in turning specific defects into efficient nonradiative centers, solving the problem of the bottleneck in SRH recombination in wide-band-gap materials.

Our analysis is bolstered by first-principles calculations performed within the framework of density functional theory (DFT) using the {\sc vasp} code \cite{VASP}. We used the Heyd-Scuseria-Ernzerhof (HSE) functional \cite{Heyd_JCP_2003} with a fraction of screened Fock exchange $\alpha$=$0.31$ to provide an accurate description of the electronic structure of GaN---a requirement for obtaining reliable results for defect levels \cite{Freysoldt_RMP_2014,Alkauskas_PRL_2008}. Interaction between ions and valence electrons were treated with the PAW approach \cite{VASP}. Wavefunctions were expanded in plane waves (using 400 eV for the kinetic energy cutoff), and the Brillouin zone was sampled at $k=(1/4, 1/4, 1/4)$. Formation energies and charge-state transition levels of defects \cite{Freysoldt_RMP_2014} were calculated using 96-atom supercells. The Freysoldt correction was applied for charged systems \cite{Freysoldt_PRL_2009}.

We consider nonradiative processes that occur via multiphonon emission \cite{Abakumov}. We calculate nonradiative capture coefficients $C^{nr}_{\{n,p\}}$ (units: cm$^3$s$^{-1}$) within the static approach \cite{Paessler_pssb_1975}, using an effective one-dimensional approximation for phonons \cite{Alkauskas_PRB_2014,Alkauskas_PRL_2012} (for alternative formulations, see Refs.~\onlinecite{Shi_PRB_2015,Barmparis_PRB_2015}). The special phonon mode is not an eigenmode of the system, but it represents all vibrations that couple to the change of defect's geometry caused by carrier capture. Capture coefficients are given by \cite{Alkauskas_PRB_2014}:
\begin{eqnarray}
\begin{aligned}
C_{\{n,p\}}&(T) =  Vf\eta_{sp} g\frac{2\pi}{\hbar}W_{if}^2\sum_{m,n} w_m (T) \label{capt} \\
& \times \left| \left \langle\chi_{im}|Q+\Delta Q|\chi_{fn}\right \rangle \right |^2 \delta (\Delta E + m\hbar\Omega_{i}-n\hbar\Omega_{f}).
\end{aligned}
\end{eqnarray}
$V$ is the supercell volume, $f$ is the scaling (Sommerfeld) factor needed to describe capture by charged defects \cite{Paessler_pssb_1975,Alkauskas_PRB_2014}, $g$ is the degeneracy of the final state, $W_{if}$ is the electron-phonon coupling matrix element \cite{Alkauskas_PRB_2014}. $\eta_{sp}$ accounts for spin-selection rules: $\eta_{sp}=1$ when the initial state is spin-singlet and the final state is spin-doublet, $\eta_{sp}=1/2$ when the initial state is spin-doublet and the final is spin-singlet, etc.
$\Delta E$ is the energy difference between the two states, $\Omega_{\{i,f\}}$ are effective vibrational frequencies in the initial and the final state, and $Q$ is the effective one-dimensional phonon coordinate, with $Q=0$ corresponding to the equilibrium geometry of the initial state.  Potential energy minima in the two electronic states are offset by $\Delta Q$ \cite{Alkauskas_PRB_2014,Alkauskas_PRL_2012}, as shown in Fig.~\ref{SRH}(b). The sum runs over vibrational states in the excited ($\chi_{im}$) and ground ($\chi_{fn}$) electronic states. $w_m(T)$ is the thermal occupation factor. $\delta$-functions in the sum are replaced by Gaussians with widths $\sigma=0.8\hbar\Omega_f$ \cite{Alkauskas_PRB_2014}.
Matrix elements $W_{if}$ are calculated as in Ref.~ \onlinecite{Alkauskas_PRB_2014}.

Cation vacancies have been flagged as important defects in nitrides, and invoked as nonradiative recombination centers \cite{Chichibu_APL_2005}. However, isolated gallium vacancies ($V_{\text{Ga}}$) have high formation energies and are thus unlikely to form; complexing with donor impurities such as oxygen significantly lowers the formation energy \cite{Neugebauer_APL_1996}. Indeed, mobility \cite{Kyle_JAP_2014} and positron annihilation \cite{Nykaen_APL_2012} studies indicate that there are at least $N=10^{16}$ cm$^{-3}$ gallium-vacancy related defects in GaN. We will use this number as a conservative estimate for the defect concentration. However, as mentioned in the introduction, the precipitous decrease in capture rate with increasing transition energy seemingly renders these defects ineffective SRH centers in materials with larger band gaps.

\begin{figure}
\includegraphics[width=8.0cm]{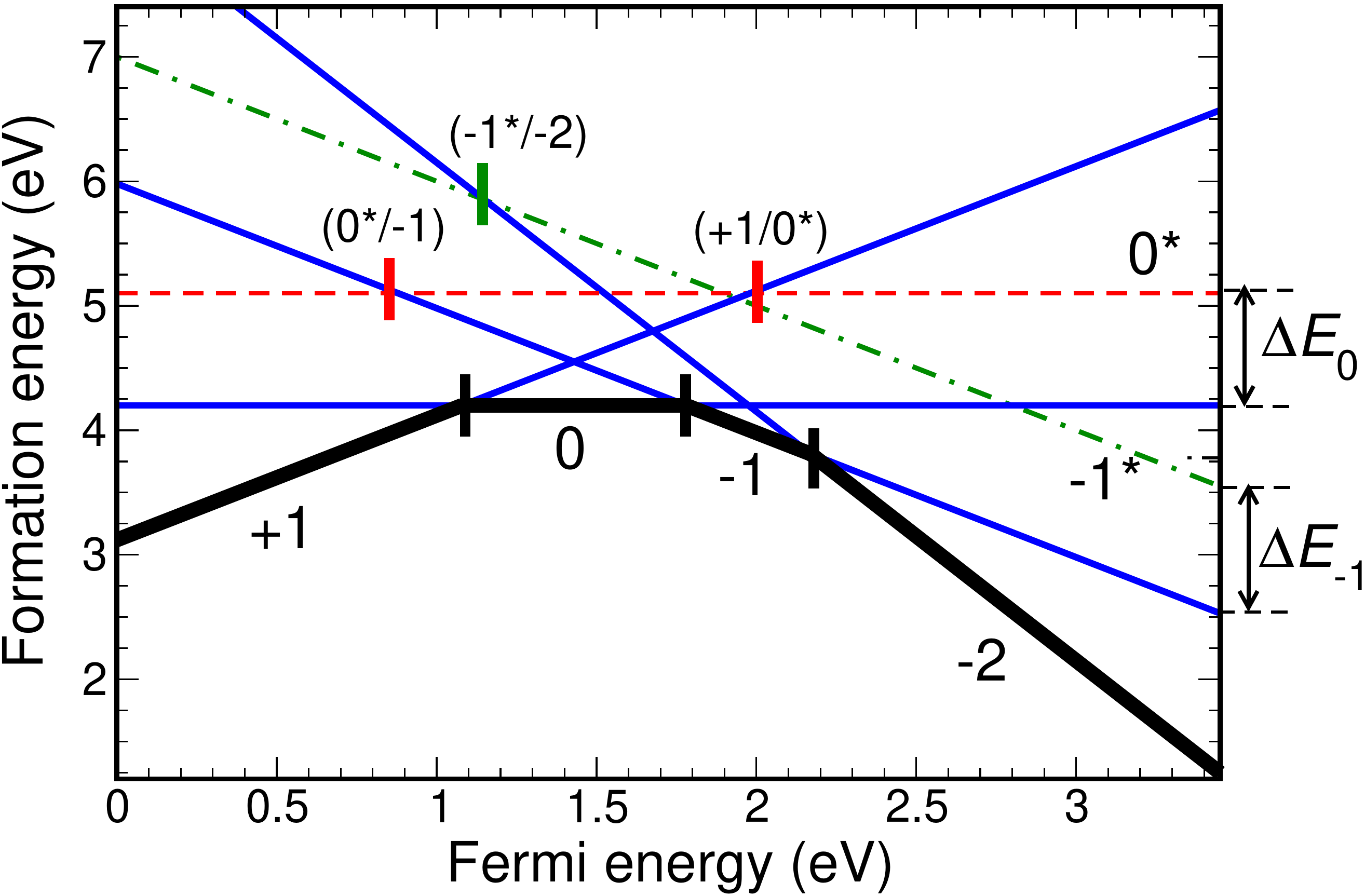} \\
\caption{(Color online) Formation energy of the $V_{\text{Ga}}$-O$_{\text{N}}$ complex in bulk GaN in different charge states as a function of Fermi energy, under Ga-rich conditions. Blue solid lines correspond to the lowest-energy configuration for each charge state; the black lines highlight the charge state with the lowest ground-state energy for a specific Fermi-level position. Red (dashed) and green (dash-dotted) lines correspond to excited states of the neutral and negatively charged state, respectively. $\Delta E_0$ and $\Delta E_{-1}$ are intra-defect excitation energies for the neutral and the negatively charged defect. Charge-state transition levels that involve excited states and that are important for SRH recombination are explicitly marked.
}
\label{vgao}
\end{figure}

We now demonstrate that capture into electronic excited states can provide an extremely efficient recombination channel. To illustrate this mechanism we will focus on a specific complex, $V_{\text{Ga}}$-O$_{\text{N}}$, which we will refer to as the VO center. All subsequent discussion also applies to the $V_{\text{Ga}}$-H defect which has a very similar electronic structure to that of VO \cite{Supplemental}. However, this defect has a larger formation energy and is less likely to form than the VO center. The calculated formation energies \cite{Lyons_pssb_2015} for the different charge states of the VO center in GaN  are shown in Fig.~\ref{vgao}. The $(+/0)$, $(0/-)$, and $(-/2-)$ charge-state transition levels occur at 1.08, 1.78, and 2.17 eV above the valence-band maximum (VBM), respectively.

For a defect with a single energy level in the gap (Fig.~\ref{SRH}) and assuming that carrier re-emission is negligible (entirely justified for deep levels in wide-band-gap semiconductors), the SRH recombination rate is \cite{Shockley_PR_1952,Hall_PR_1952}
\begin{equation}
R=N\frac{C_nC_pnp}{C_nn + C_pp}.
\label{SRH2}
\end{equation}
where $N^{0}+N^{-}=N$ is the total defect density.
If photo-generated or injected carrier densities are much larger than the background carrier density, then $n\approx p$ and $R$ can be written as $R=An$, where $A$ (units: s$^{-1}$) is the SRH coefficient:
\begin{equation}
A=N C_{tot}; \text{ } C_{tot}=\frac{C_n C_p}{C_n + C_p}.
\label{A}
\end{equation}
This equation makes clear that the {\it slower} of the two processes determines the overall rate. For a defect such as the VO center [Fig.~\ref{vgao}] with multiple levels in the gap, one might think that the closer positioning of the (+/0) and ($-$1/$-$2) transition levels to the respective band edges would be an advantage; in reality, such additional transition levels reduce the efficiency of SRH recombination. For the defect with multiple charge states one can derive \cite{Sah_PR_1958,Supplemental} a SRH coefficient $A=NC_{tot}$, where
\begin{equation}
\begin{aligned}
\label{multi}
C_{tot} &= \frac{1}{G}\left(C^{+1}_{n}+\frac{C^{+1}_{n}C^{0}_{n}}{C^{0}_{p}}+\frac{C^{+1}_{n}C^{0}_{n}C^{-1}_n}{C^{0}_{p}C^{-1}_{p}}\right);
\\
G &= 1+\frac{C^{+1}_n}{C^0_p} + \frac{C^{+1}_nC^{0}_n}{C^{0}_pC^{-1}_p} + \frac{C^{+1}_nC^{0}_nC^{-1}_n}{C^{0}_pC^{-1}_pC^{-2}_p}.
\end{aligned}
\end{equation}
The superscripts indicate the charge state, the subscripts specify the carrier type. We exclude the consideration of radiative capture processes, which are characterized by coefficients $C_{rad}=10^{-14}-10^{-13}$ cm$^{3}$s$^{-1}$ [Ref. ~\onlinecite{Reshchikov_JAP_2005}].  Assuming $N=10^{16}$ cm$^{-3}$ this yields $A$ coefficients three-four orders of magnitude smaller than the values of $A\approx10^7$ s$^{-1}$ experimentally determined for InGaN \cite{David_APL_2010}.

We calculated \cite{Supplemental} all capture coefficients appearing in Eq.~(\ref{multi}). We assumed $T$=120$^{\text{o}}$C, a typical internal temperature of operating LEDs \cite{Pan_APEX_2012}. Explicit calculations for InGaN alloys are computationally prohibitive.  Based on calculations at select alloy compositions we have found that the variation in defect properties is predominantly determined by the change in the lattice parameters between GaN and InGaN; explicit interactions with In atoms affect defect levels by less than 0.1 eV. We therefore determined charge-state transition levels of the VO defect in GaN supercells with lattice parameters expanded to match the volume of InGaN with 5, 10, and 20\% In. The resulting transition levels are then positioned within the InGaN band gap by (i) aligning the average of the top three valence bands in the expanded cell to that of unstrained GaN using the absolute valence-band deformation potential of GaN \cite{VdW_APL_1997}; (ii) aligning band edges of unstrained GaN and unstrained InGaN using band offsets from Ref.~\onlinecite{Moses_APL_2010}. The dependence of transition levels on band gap is given in Ref.~\onlinecite{Supplemental}.

Our calculations reveal that the capture coefficient $C_{tot}$ [Eq.~(\ref{multi})] is smaller than $10^{-19}$ cm$^{3}$s$^{-1}$ for band gaps $2.20-3.45$ eV. The reason for these negligibly small coefficients is that the defect gets stuck in the ``extreme'' charge states $+1$ and $-2$. This follows from Eq.~(\ref{multi}) in the following way \cite{Sah_PR_1958,Supplemental}: the total recombination coefficient $C_{tot}$ is a sum of coefficients corresponding to pairs of charge states with charges differing by one. For each pair the recombination rate is given by Eq.~(\ref{SRH2}), with $N$ now representing the concentration of defects in those particular two charge states. \cite{Sah_PR_1958} Let us take the $(+1,0)$ pair of charge states as an example. In steady state, the number of defects in each of the charge states is constant, and detailed balance gives $N^{+1}C^{+1}_n=N^{0}C^0_p$, and thus $N^{+1}/N^{0}=C^0_p/C^{+1}_n$. The $(+1/0)$ charge-state transition level is much closer to the VBM than to the CBM. Consequently, $C^0_p\gg C^{+1}_n$, and therefore $N^{+1}\gg N^{0}$. Similar reasoning for the ($-$1,$-$2) pair leads to $N^{-2}\gg N^{-1}$. Thus almost all defects are in either the $+1$ or $-2$ charge states. The $+1$ state captures electrons very inefficiently, while the $-2$ state captures holes very inefficiently, leading to the overall low SRH recombination rate.

\begin{figure}
\includegraphics[width=8.0cm]{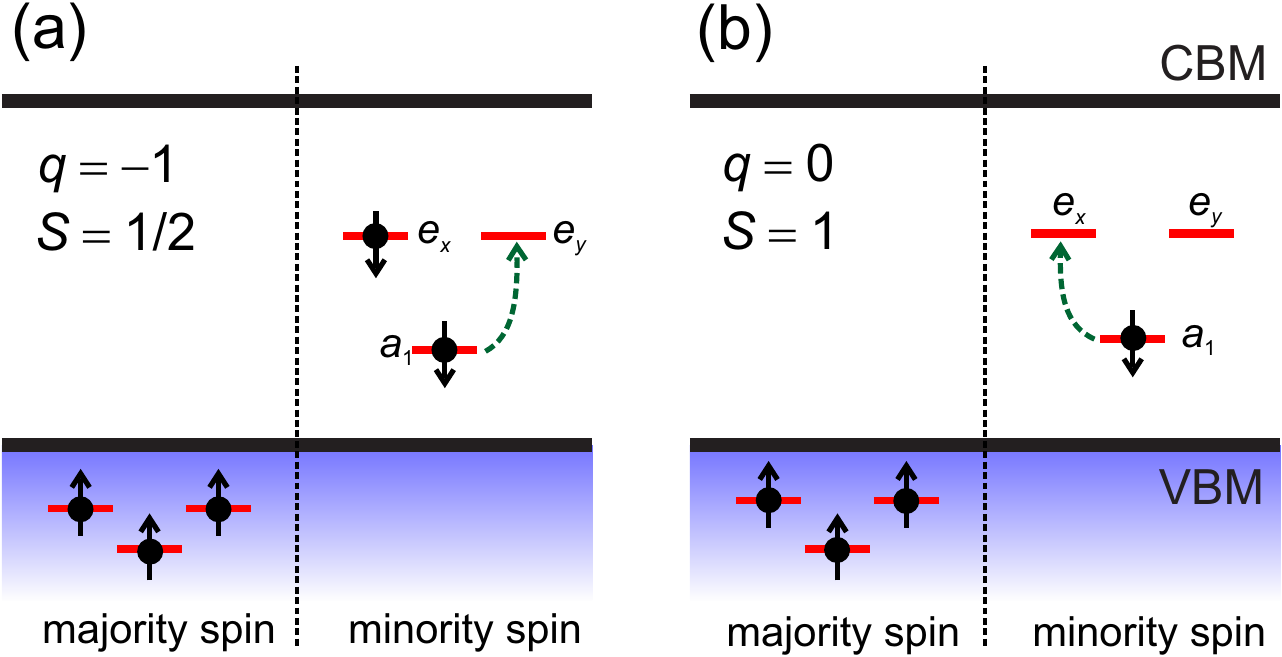}
\caption{(Color online) Single-particle states of the VO defect for charge states $q=-1$ (a) and $q=0$ (b). Intra-defect excitation occurs by promoting an electron from the $a_1$ state to the $e$ state.
}
\label{exc}
\end{figure}

Our considerations so far did not take {\it excited} electronic states into account. We will show that such states exist for the $-$1 and 0 charge states. Single-particle defect states at the VO center arise from the interaction of dangling bonds on three N atoms. The lower-lying state results from a symmetric combination, while the two higher-lying states result from an anti-symmetric combination. Were the symmetry of the center $C_{3v}$, these single-particle states would belong to $a_1$ and $e$ irreducible representations; we will retain these labels also for lower symmetries. In charge state $q$=$-$2 all levels are filled, and no excited states are possible. For $q$=$-$1 (spin $S$=1/2) the electronic configuration is $a^1_1e^2$ in the majority-spin channel, and $a^1_1e^1$ in the minority-spin channel. The excited state
is an excitation from the $a_1$ state to the $e$ state, as shown in Fig.~\ref{exc}(a). For charge state $q$=0 ($S$=1) the electronic configuration is $a^1_1e^0$ in the minority-spin channel, and again, an excitation can occur by promoting the $a_1$ electron to the $e$ state [Fig.~\ref{exc}(b)]. In the ground state of the $q$=+1 charge state ($S$=3/2) all defect states are filled in the majority-spin channel, and empty in the minority-spin channel, and thus there are no spin-conserving excited states.

DFT is a ground-state theory, and in order to describe excited states one has to go beyond DFT. One way to do this is to approximate total energy differences by differences in single-particle Kohn-Sham eigenvalues in the spirit of the generalized Koopmans theorem \cite{Lany_PRB_2009,Dabo_PRB_2010,Deak_PRB_2010}. For transitions for which actual comparison could be made we have explicitly verified  \cite{Supplemental} that this approach is very accurate for the VO defect when calculations are performed with the HSE hybrid functional. By mapping ground-state and excitation energies along the relevant configuration coordinates \cite{Ma_PRB_2010} we can determine potential energy surfaces in the excited states \cite{Supplemental}. We obtain intra-defect excitation energies $\Delta E_{0}=0.90$ eV and $\Delta E_{-1}=1.02$ eV, as shown in Fig.~\ref{vgao}.

Excited states drastically change the dynamics of SRH recombination. Let us again consider the $(+1,0)$ pair of charge states (which will turn out to be the most important for SRH recombination at VO defects). Without excited states recombination via this pair is slow: almost all defects defects are in the $+1$ state, which captures electrons inefficiently since the $(+1/0)$ transition is far from the CBM (2.37 eV in GaN). When excited states are taken into account, capture of an electron into the $0^*$ state can occur (the phrase ``capture into'' refers to the {\it final} state of the process). The $(+1/0^*)$ transition [vertical red bar in Fig.~\ref{vgao})] is closer to the CBM by an amount $\Delta E_0$; i.e., the $(+1/0^*)$ level is 2.37$-$0.90=1.47 eV below the CBM, resulting in much more efficient electron capture. After this capture process occurs, the system quickly relaxes from $0^*$ to $0$ via intra-defect relaxation \cite{Supplemental}. Once in the $0$ charge state, the defect captures a hole very efficiently, closing the recombination cycle.

Also important for the SRH recombination is hole capture by the $-2$ state into the $-1^*$ state. In GaN the $(-2/-1)$ charge-state transition level is 2.17 eV from the VBM, resulting in negligibly slow hole capture. The presence of the $-1^*$ state reduces this separation by $\Delta E_{-1}=1.02$ eV, resulting in a substantial increase in hole capture. Equally important is hole capture by the $-1$ state: the presence of the $0^*$ state makes this process orders of magnitude more efficient, because the $(-1/0^*)$ transition level is much closer to the VBM than the $(-1/0)$ level [Fig.~\ref{vgao}]. Calculations show that the latter two processes are responsible for removing defects from charge states $-2$ and $-1$, and most recombination proceeds via states $+1$ and $0$.

\begin{figure}
\includegraphics[width=8.5cm]{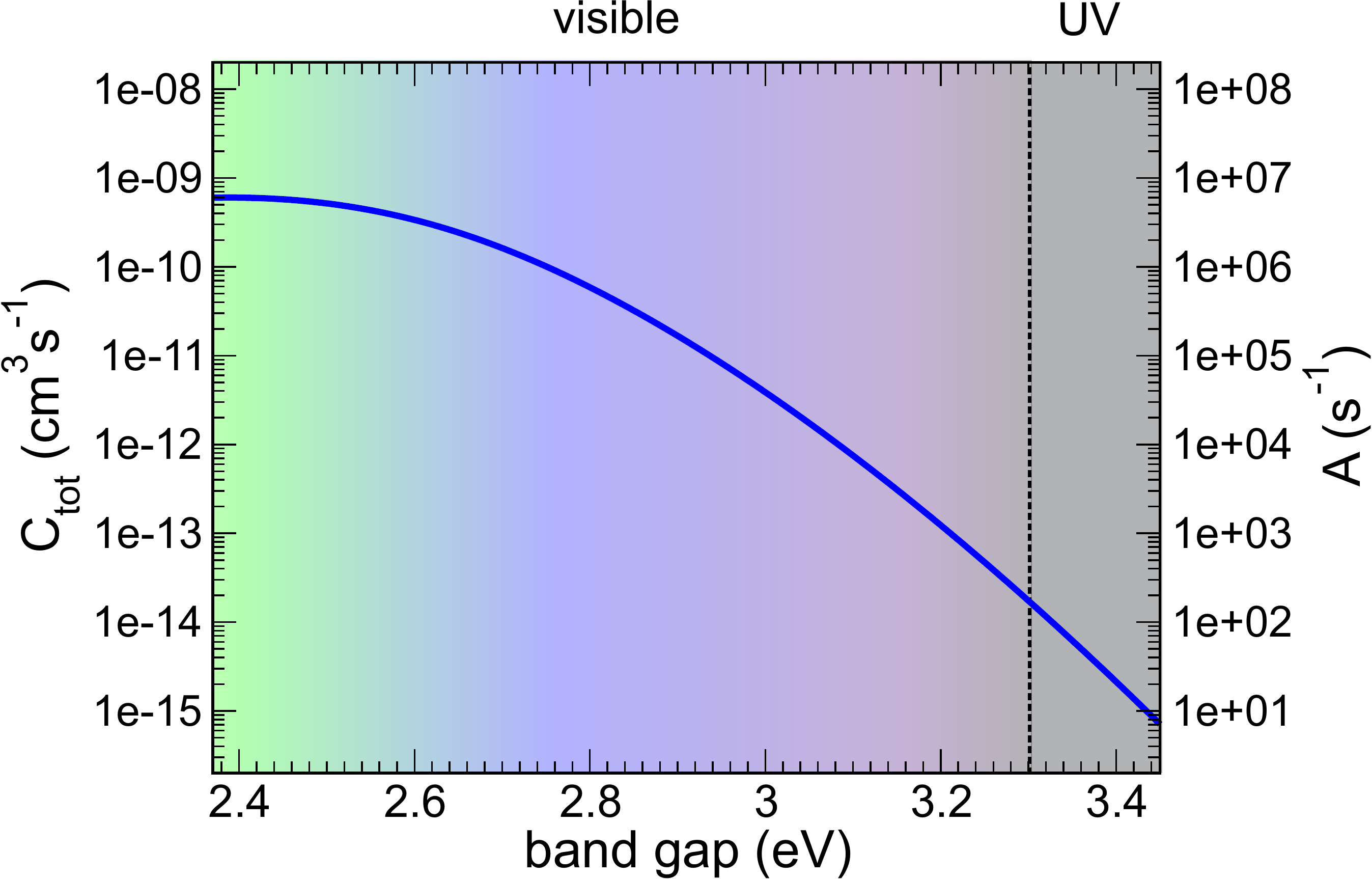} \\
\caption{(Color online) Calculated total capture coefficient $C_{tot}$ (left vertical axis) and SRH coefficient $A=NC_{tot}$ (right vertical axis) as a function of band gap for gallium-vacancy complexes with oxygen. We assumed a defect concentration $N=10^{16}$ cm$^{-3}$.}
\label{capture}
\end{figure}

Analysis of Fig.~\ref{vgao} would seem to suggest that electron capture by the $0$ charge state (configuration $a_1^1e^0$) into the $(-1^*)$ charge state (configuration $a_1^0e^2$) should also be beneficial for SRH recombination. However, this is actually an Auger process, because it requires that one electron is captured from the conduction band to the $e$ state, while the other at the same time is promoted from $a_1$ to $e$. Such four-state processes are expected to be slower \cite{Abakumov} and are not taken into account in our analysis.

Quantitatively, the total capture coefficient in the presence of excited states is given by an equation analogous to Eq.~(\ref{multi}), with $C_n^{+1}$, $C_p^{-1}$, and $C_p^{-2}$ replaced by coefficients that describe capture into the respective excited states. The overall capture coefficient is shown in Fig.~\ref{capture}. $C_{tot}$ is equal to $1\times10^{-10}$ cm$^3$s$^{-1}$ for InGaN alloys emitting in the blue (band gap 2.75 eV) and $6\times10^{-10}$ cm$^3$s$^{-1}$ for InGaN emitting in the green (2.40 eV). For defect concentration $N=10^{16}$ cm$^{-3}$, this gives SRH coefficients $A=NC_{tot}=1\times10^{6}$ s$^{-1}$ and $A=6\times10^{6}$ s$^{-1}$, respectively. For the blue InGaN the determined $A$ coefficient is very close to the ones found in actual LEDs \cite{David_APL_2010}.

We are now in a position to determine the impact of SRH recombination on efficiency of LEDs. At low injected carrier densities, when third-order processes can be ignored, the internal quantum efficiency is given by $\eta=Bn/(A+Bn)$. We take $n=10^{18}$ cm$^{-3}$, a typical carrier density in operating LEDs \cite{David_APL_2010}. Since $B \approx 4\times10^{-11}$ cm$^{3}$s$^{-1}$ \cite{Kioupakis_NJP_2013}, we find that $N=10^{16}$ cm$^{-3}$ VO centers would decrease the quantum efficiency of blue LEDs by $\sim$2.5\% and green LEDs by 13\%. A defect concentration of $N=10^{17}$ cm$^{-3}$ would decrease efficiencies of blue LEDs by 20\%, and those of green LEDs by as much as 60\%. Note that in the latter case, where most of the recombination is nonradiative, the presence of defects modifies the charge-neutrality condition and we can no longer take $n=p$, an assumption used in deriving Eqs.~(\ref{A}) and (\ref{multi}). Still, these equations provide a reasonable estimate of the SRH coefficient. Our analysis shows that gallium-vacancy complexes are an important source of efficiency loss in nitride light emitters.

In conclusion, we have found that electronic excited states can play a crucial role in Shockley-Read-Hall recombination in wide-band-gap semiconductors. Our example for gallium vacancy complexes in nitride semiconductors demonstrated that inclusion of excited states enhances nonradiative recombination rates by many orders of magnitude. We suggest that such mechanisms should be ubiquitous in wide-band-gap materials.   Excited states similar to the ones considered in this work can occur at cation vacancies and their complexes not only in nitrides, but also in oxides as well as carbide materials. More generally, excited states of defects can also affect recombination mechanisms other than the multiphonon emission process considered in this work. For example, in the so-called impurity Auger process \cite{Abakumov} the recombination rates also decrease exponentially when the band gap of the material increases, seemingly making such processes unimportant for wide-band-gap materials. Inclusion of excited states of defects could drastically change the impact of this mechanism as well.

\begin{acknowledgments}
We acknowledge M. W. Doherty, M. A. Reshchikov, J. S. Speck, C. Weisbuch and Q. Yan for fruitful interactions. A.\ A.\ was supported by Marie Sk{\l}odowska-Curie Action of the European Union (project N\textsc{itride}-SRH, grant No.~657054). Work at UCSB was supported by the U.\ S.\ Department of Energy (DOE), Office of Science, Basic Energy Sciences (BES) under Award No.~$\text{DE-SC}$0010689. Computational resources were provided by the National Energy Research Scientific Computing Center, which is supported by the DOE Office of Science under Contract No.~DE-AC02-05CH11231.
\end{acknowledgments}

\pagebreak
\widetext
\begin{center}
\textbf{\large Role of excited states in Shockley-Read-Hall recombination in wide-band-gap semiconductors: Supplemental Material}

\end{center}
\setcounter{equation}{0}
\setcounter{figure}{0}
\setcounter{table}{0}
\setcounter{page}{1}
\makeatletter
\renewcommand{\theequation}{S\arabic{equation}}
\renewcommand{\thefigure}{S\arabic{figure}}
\renewcommand{\bibnumfmt}[1]{[S#1]}
\renewcommand{\citenumfont}[1]{S#1}

\section{Formation energy of the $V_{\text{Ga}}-$H complex}
\label{sec:vgah}

\begin{figure*}[h]
\includegraphics[width=8.5cm]{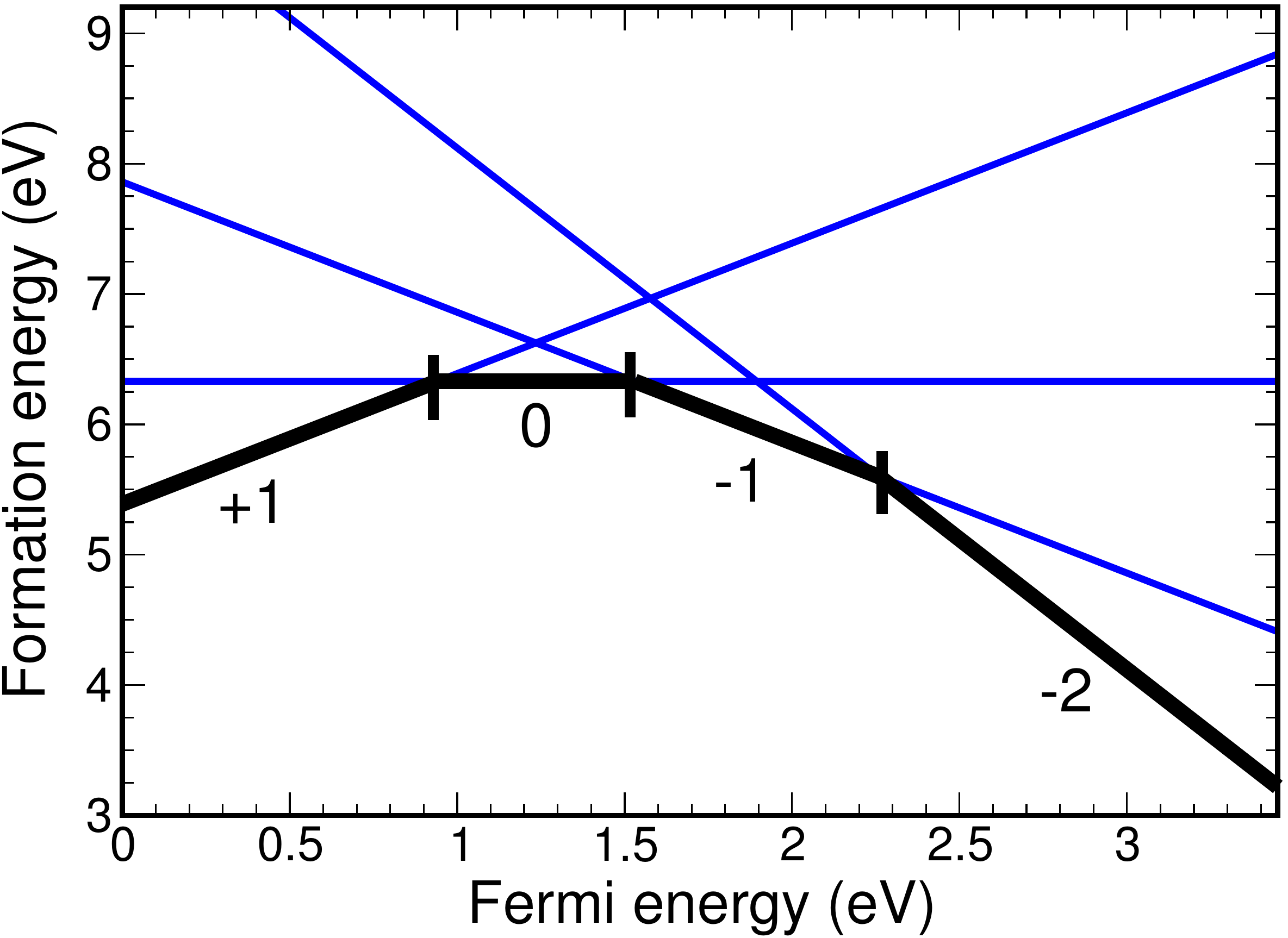}
\caption{Formation energy of the $V_{\text{Ga}}-$H complex in bulk GaN in different charge states as a function of Fermi energy, under Ga-rich conditions. Blue solid lines correspond to the lowest-energy configuration for each charge state; the black lines highlight the charge state with the lowest ground-state energy for a specific Fermi-level position.}
\label{vgah}
\end{figure*}

The formation energy of the $V_{\text{Ga}}-$H complex is shown in Fig.~\ref{vgah}.

\section{Capture coefficients for a defect with multiple charge states}

Carrier recombination at defects with many charge states has been considered by Sah and Shockley \cite{Sah_PR_1958}. Let us take the VO defect with charge states $+1$, 0, $-1$, and $-2$ [Fig.~2 of the main text]. Under steady-state conditions the concentration of defects in each charge state is constant. Thus:
$C^{+1}_nN^{+1}=C^{0}_pN^{0}$, $C^{0}_nN^{0}=C^{-1}_pN^{-1}$, $C^{-1}_nN^{-1}=C^{-2}_pN^{-2}$ (we assume $n=p$). Using the normalization condition $N^{+1}+N^{0}+N^{-1}+N^{-2}=N$ we can find the expressions for all concentrations in terms of $N$:
\begin{eqnarray}
N^{+1}=\frac{N}{G};\nonumber \\
N^{0}=\frac{C_n^{+1}}{C_p^0}\times\frac{N}{G};\nonumber\\
N^{-1}=\frac{C_n^{+1}C_n^{0}}{C_p^0C_p^{-1}}\times\frac{N}{G};\nonumber\\
N^{-2}=\frac{C_n^{+1}C_n^{0}C_n^{-1}}{C_p^0C_p^{-1}C_p^{-2}}\times\frac{N}{G};\nonumber\\
G=1+\frac{C^{+1}_n}{C^0_p} + \frac{C^{+1}_nC^{0}_n}{C^{0}_pC^{-1}_p} + \frac{C^{+1}_nC^{0}_nC^{-1}_n}{C^{0}_pC^{-1}_pC^{-2}_p}.
\label{N}
\end{eqnarray}
 The total recombination rate is given by the total rate of electron or hole consumption. For example, it can be written as:
\begin{equation}
r=(N^{+1}C_n^{+1}+N^{0}C_n^{0}+N^{-1}C_n^{-1})n.
\label{rate2}
\end{equation}
Using the expressions for $N$'s from Eq.~(\ref{N}) and keeping in mind the definition of $C_{tot}$ via $r=NC_{tot}n$, we arrive at Eq.~(4) of the main text.

\section{Parameters of calculations of nonradiative capture rates \label{Par}}

In this Section we provide a summary of parameters used in calculations of nonradiative carrier capture coefficients at the $V_{\text{Ga}}-$O$_{\text{N}}$ (VO) defect. Coefficients themselves have been calculed using Eq.~(1) of the main text. All parameters, except for $\eta_{sp}$, were discussed in Refs.~\onlinecite{Alkauskas_PRL_2012,Alkauskas_PRB_2014}. $\eta_{sp}$ is a spin factor that describes spin selection rules. E.g., for carrier capture from an initial spin-singlet state to a final spin-doublet state, $\eta_{sp}=1$ (regardless of the initial state of the defect, carriers of both spins can be captured); for capture from an initial doublet state to a final spin-singlet state, $\eta_{sp}=1/2$ (regardless of the initial state of the defect, only carriers having a spin opposite to that of the defect can be captured), etc.
In Ref.~\onlinecite{Alkauskas_PRB_2014} only transitions with $\eta_{sp}=1$ were considered.

For the Sommerfeld factor $f$ we used analytical expressions from Ref.~\onlinecite{Paessler_pssb_1976}:
\begin{eqnarray}
f=\frac{2\pi |Z|}{a_{B}k_T}, \hspace{2mm} Z>0; \nonumber \\
f=\frac{2\pi |Z|}{a_{B}k_T}\exp\left(-\frac{2\pi |Z|}{a_{B}k_T}\right), \hspace{2mm} Z<0. \\
\end{eqnarray}
Here $k_T$ is thermal momentum corresponding to $t=120^{\text{o}}$C, $a_B$ is the Bohr radius in the material (different for electrons and holes because of differences in effective masses $m_e$ and $m_e$), $Z>0$ corresponds to attractive centers, $Z<0$ to repulsive centers.

The mass-weighted distortion parameter $\Delta Q$ is given by:
\begin{equation}
\left(\Delta Q\right)^2 = \sum_{\alpha, t} m_{\alpha} \Delta R_{\alpha t}^2,
\label{vib2}
\end{equation}
where the sum runs over all atoms $\alpha$; $t=\{x,y,z\}$. $\Delta R_{\alpha t}$ describes the change of the defect geometry associated with the change of its electronic state.

Parameters to calculate nonradiative capture coefficients for all elementary processes are listed in Table \ref{Table1}. For excited states, potential energy surfaces have been determined as explained in Sec.~\ref{Exc}. Electron-phonon matrix elements $W_{if}$ have been calculated as described in Ref.~\onlinecite{Alkauskas_PRB_2014}. We get $W_{if}=2.9\times10^{-2}$ eV/(amu$^{1/2}$\AA) for the coupling between the valence-band maximum (VBM) and the $a_1$ state, $W_{if}=5.0\times10^{-2}$ eV/(amu$^{1/2}$\AA) for the coupling between the VBM and $e$ states, $W_{if}=5.1\times10^{-3}$ eV/(amu$^{1/2}$\AA) for the coupling between the conduction-band minimum (CBM) and the $a_1$ state, $W_{if}=1.7\times10^{-3}$ eV/(amu$^{1/2}$\AA) for the coupling between the CBM and the $e$ states. These values have been determined from the calculation for the $0$ charge state in 96-atom supercells.

\begin{table*}[b]
\caption{Parameters entering the calculation of nonradiative carrier capture coefficients: total mass-weighted distortion $\Delta Q$, energy difference $\Delta E$ in GaN, energies of effective vibrations $\hbar\Omega_{\{i,f\}}$ in the initial and the final state, degeneracy factor $g$ of the final state, spin factor $\eta_{sp}$, and the Sommerfeld scaling factor $f$.}

\begin{ruledtabular}
\begin{tabular}{l c c c c c c c c}

& & $\Delta Q$ & $\Delta E$ & $\hbar\Omega_{i}$ & $\hbar\Omega_{f}$ & $g$ & $\eta_{sp}$ & $f$ \\
transition & carriers  & amu$^{1/2}$\AA & eV      & meV         & meV           &   &&      \\
\hline
$(+/0)$    &electrons  & 2.03  & 2.38 & 33.8  & 32.6  & 3  & 1/2  & 5.25                    \\
           &holes      & 2.03  & 1.07 & 32.6  & 33.8  & 1  & 1/2  & 1                       \\
\hline
$(0/-)$    &electrons  & 1.85  & 1.67 & 35.1  & 36.9  & 2 & 1/2   & 1                       \\
           &holes      & 1.85  & 1.78 & 36.9  & 35.1  & 2 & 1/2   & 10.5                    \\
\hline
$(-/-$$2)$ &electrons  & 1.77  & 1.28 & 36.5  & 39.3  & 1 & 1/2   & $2.8\times10^{-2}$      \\
           &holes      & 1.77  & 2.17 & 39.3  & 36.5  & 3 & 1     & 21                      \\
\hline
\hline
$(+/0^{*})$&electrons  & 1.47  & 1.47 & 33.8  & 36.2  & 2 & 1/2   & 5.25                    \\
\hline
$(0^*/-)$  &holes      & 1.70  & 0.88 & 36.7  & 36.2  & 2 & 1/2   & 10.5                    \\
\hline
$(-$$1^*/-$$2)$&holes  & 0.93  & 1.15 & 39.3  & 45.4  & 1 & 1     & 21                      \\
\end{tabular}
\label{Table1}
\end{ruledtabular}
\end{table*}

\section{Charge-state transition levels of the $V_{\text{Ga}}-$O$_{\text{N}}$ complex in InGaN alloys \label{InGaN}}

\begin{figure*}[h]
\includegraphics[width=8.5cm]{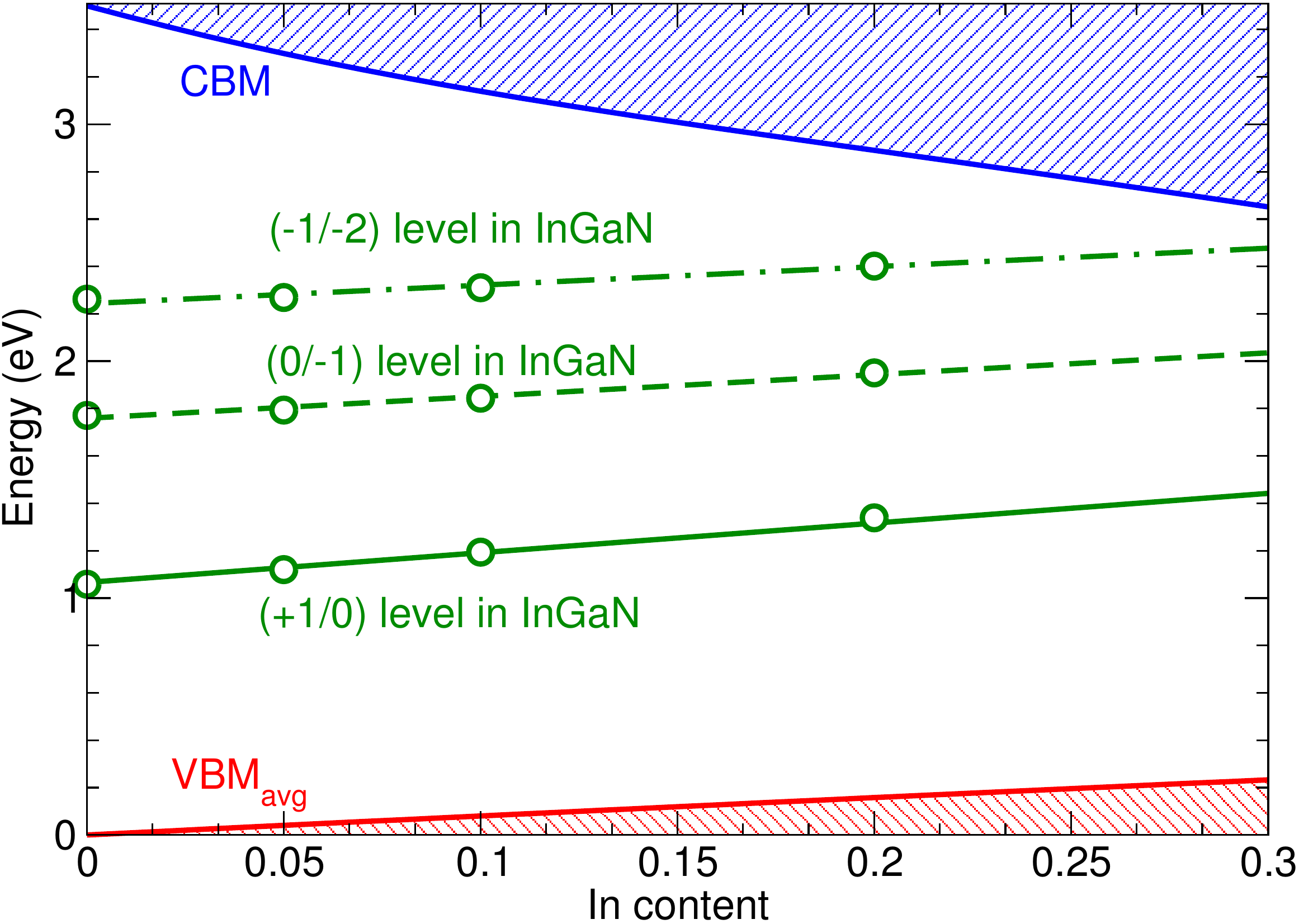}
\caption{Position of charge-state transition levels of the $V_{\text{Ga}}-$O$_{\text{N}}$ complex within the band gap of InGaN as a function of In content.
For the valence band, an average over the top three valence bands is shown \cite{Moses_APL_2010,Moses_JCP_2011}.
VBM$_{\rm avg}$ of bulk GaN was chosen as the reference energy.
The green dot-dashed line takes the change in volume of the InGaN alloy into account, interpolating between the explicit calculations given by the green circles.}
\label{vgao-inagan}
\end{figure*}

Charge-state transition levels of $V_{\text{Ga}}-$O$_{\text{N}}$ complex in In$_{x}$Ga$_{1-x}$N alloys are shown in Fig.~\ref{vgao-inagan}.
Calculations have been performed as described in the main text for $x=$ 0.05, 0.1, and 0.2. The dependence of the VBM, CBM, and charge-state transition levels on $x$ have been fitted to analytical expressions to produce Fig. 4 of the main text.

\section{Excited states of the $(V_{\text{Ga}}-$O$_{\text{N}})$ defect \label{Exc}}

\subsection{Methodology}
\label{sec:meth}

Density functional theory (DFT) is a ground-state theory, and one has to go beyond it to describe excited states. It is well known that when using DFT with approximate functionals, single-particle KS eigenvalues are not in general equal to electron removal and addition energies \cite{Mori-Sanchez_PRL_2008}. In local and semi-local approximations to DFT this happens because of self-interaction errors inherent to these functionals \cite{Mori-Sanchez_PRL_2008}. A lot of research efforts (see, e.g., \cite{Lany_PRB_2009,Dabo_PRB_2010}) have been aimed at developing approximate functionals that would cure this shortcoming and would ensure the fulfillment of the condition $\text{EA} = E(N) - E(N+1) = - \varepsilon_i(N)$, where EA is the electron affinity, $E(N)$ is the total energy of a system with $N$ electrons, and $\varepsilon_i(N)$ is the eigenvalue of the state to which the additional electron is added; the number in parentheses indicates the total number of electrons in the system. Similarly, one has $\text{IP} = E(N-1) - E(N) = - \varepsilon_i(N)$, where IP  stands for the ionization potential, and $\varepsilon_i(N)$ is now the eigenvalue of the state from which the electron is removed. The two equalities above specify the so-called generalized Koopmans condition for density functionals \cite{Lany_PRB_2009,Dabo_PRB_2010}.

In comparison to local and semi-local functionals, self-interaction errors are greatly reduced in hybrid functionals. More specificaly, it has been demonstrated that in hybrid functional calculations (using the HSE functional) of defects in group-IV semiconductors the Koopmans condition is fullfilled to a high degree \cite{Deak_PRB_2010}. In this work we have excplicitly verified this for the VO defect: when calculations are performed in HSE hybrid functional with a fraction $a=0.31$ (which reproduces the band gap of GaN), single-particle eigenvalues and total energy differences agree to within $0.1$ eV. Note that when such comparison is made it is important to include finite-size supercell corrections \cite{Freysoldt_pssb_2011} not only on total energies, but also for single-particle eigenvalues \cite{Lany_PRB_2008,Chen_PRB_2013}.

The accuracy of substituting total energy differences with eigenvalues can be explicitly verified for the highest occupied and the lowest unoccupied defect states. This allows us to prove that this is a very good approximation when the electron is added to or removed from either the $a_1$ state (calculations for charge state $q=+1$ and $q=0$, respectively) and $e$ states ($q=0,-1$ and $q=-1,-2$). We are therefore ensured that this would remain true even if an electron were removed from the state one below the highest-occupied state (e.g., state $a_1$ of the negatively charged defect) or added to the state one above the lowest unoccupied state (e.g., state $e$ of the positively charged defect).

\subsection{Neutral charge state}

\begin{figure*}[h]
\includegraphics[width=14cm]{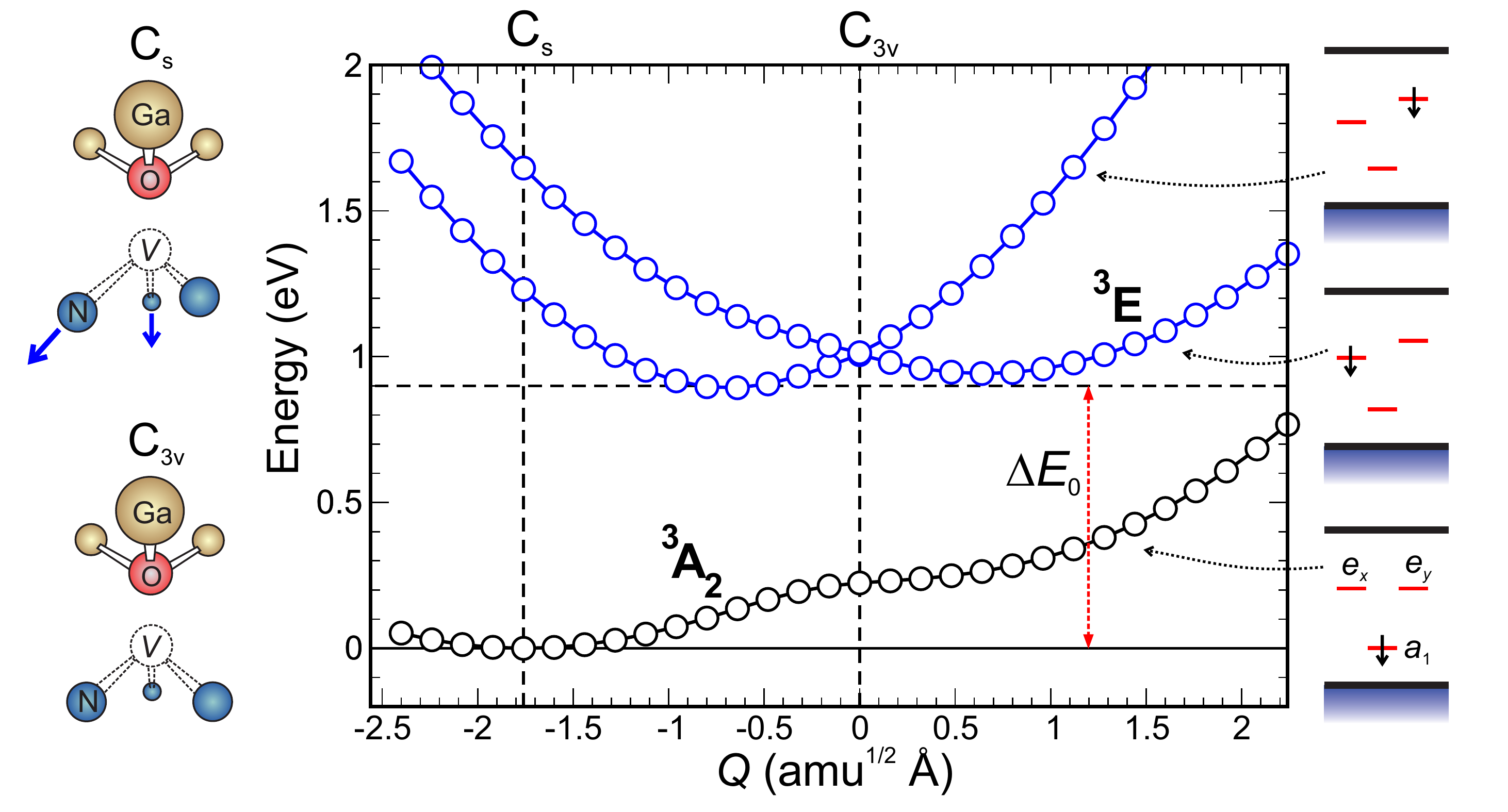}
\caption{One-dimensional configuration coordinate diagram of the VO defect in the neutral charge state. The configuration coordinate interpolates the defect geometry in the distorted geometry 
and a $C_{3v}$ geometry
The electronic states are labeled according to the irreducible representations of the $C_{3v}$ point group. Ball-and-stick models on the left show geometries of the defect in these two configurations. The position of the vacancy is indicated with a ``$V$''.  Solid arrows indicate relaxations of N atoms. Figures on the right show single-particle states and occupations in the minority spin channel for each of the electronic states. These should be read as follows. In the $^3A_2$ electronic state the $a_1$ defect state is filled. The $^3E$ electronic state with lower energy always has the lower $e$ state filled, while the one with the higher energy has the higher $e$ state filled. The latter two are degenerate for the $C_{3v}$ symmetry.
}
\label{ccd0}
\end{figure*}

The ground state of the neutral center is a spin triplet. In the spin-majority channel all three Kohn-Sham (KS) states associated with the defect are filled, while in the spin minority channel the electronic configuration is $a_1^1e^0$ (superscript indicates the occupation of the state) (see diagrams to the side of Fig.~\ref{ccd0}). Were the defect geometry exactly of $C_{3v}$ symmetry, the total electronic wavefunction would correspond to the $^3A_2$ irreducible representation. For convenience we will retain the labeling corresponding to $C_{3v}$ symmetry also in the case of lower symmetries. The excited state corresponds to the electronic configuration $a_1^0e^1$ in the spin-minority channel. Since there are two $e$ states, there are two branches of the excited state. In $C_{3v}$ symmetry the excited state would correspond to the $^3E$ irreducible representation.

We now use the methodology described in Sec.~\ref{sec:meth} to determine excited-state properties of the VO defect in the neutral charge state. Let us take the neutral defect in some fixed geometry. Let $E_1^{q=0}$ be the ground-state energy for this geometry; in the superscript the charge state is explicitly indicated. We use the same convention also for single-particle eigenvalues. In the subscript we label eigenvalues of defect levels in the relevant spin-minority channel in order of increasing energy: $\varepsilon_1$ ($a_1$ state), $\varepsilon_2$ (lower $e$ state), $\varepsilon_3$ (higher $e$ state). The ground state of the neutral defect can be created when the electron is added to the $a_1$ level of the positively charged defect. The excited state of the neutral defect can be instead created when the electron is added to the $e$ level of the positively charged defect. Thus, for a {\it fixed} defect geometry the first excited electronic state of the neutral defect is $\varepsilon_2^{q=+1}-\varepsilon_1^{q=+1}$ higher than the ground state, while the second excited state is $\varepsilon_3^{q=+1}-\varepsilon_1^{q=+1}$ higher than the ground state. We have:
\begin{equation}
E_{2}^{q=0} = E_1^{q=0} + \varepsilon_2^{q=+1}-\varepsilon_1^{q=+1}
\label{r1}
\end{equation}
for the first excited state and
\begin{equation}
E_{3}^{q=0} = E_1^{q=0} + \varepsilon_3^{q=+1}-\varepsilon_1^{q=+1}
\label{r2}
\end{equation}
for the second excited state.

In previous paragraphs we described how to calculate energies of electronic exited states for a fixed atomic configuration. The next step is to determine how these excited state energies depend on the defect geometry  \cite{Ma_PRB_2010}. To do this, we mapped the potential energy surfaces along the configuration coordinate that linearly interpolates between two structures. One geometry corresponds to the atomic configuration in the ground state in which two N atoms experience a strong outward relaxation, as shown in Fig.~\ref{ccd0}. The symmetry here is $C_{s}$, with a mirror plane as the only symmetry operation.
The other geometry corresponds to a symmetric configuration, obtained by performing geometry relaxation of the defect when all three defect levels are occupied with $1/3$ of an electron. This is the the $C_{3v}$  geometry, with all N atoms at the same distance from the vacancy, as also shown in Fig.~\ref{ccd0}.

All terms in Eqs.~(\ref{r1}) and (\ref{r2}) were consistently calculated for the same geometry. The resulting 1D configuration coordinate diagram for the neutral defect is shown in Fig.~\ref{ccd0}. The excited state clearly displays the signatures of a $E\otimes e$ Jahn-Teller system \cite{Bersuker}: the two $^3E$ branches become degenerate for the $C_{3v}$ geometry. Along the configuration coordinate chosen in Fig.~\ref{ccd0} the minimum occurs to the left of the symmetric configuration, while the transition state between the other two equivalent minima of the $E\otimes e$ Jahn-Teller system \cite{Bersuker} is seen as a minimum that occurs to the right from the symmetric configuration in Fig.~\ref{ccd0}. The latter is only 0.04 eV above the minimum, indicating that the so-called quadratic Jahn-Teller terms are small \cite{Bersuker}. The results in Fig.~\ref{ccd0} show that the energy difference between $^3E$ states and the $^3A_2$ state is $\Delta E_0=0.90$ eV.

\subsection{Negative charge state}

\begin{figure*}[b]
\includegraphics[width=14cm]{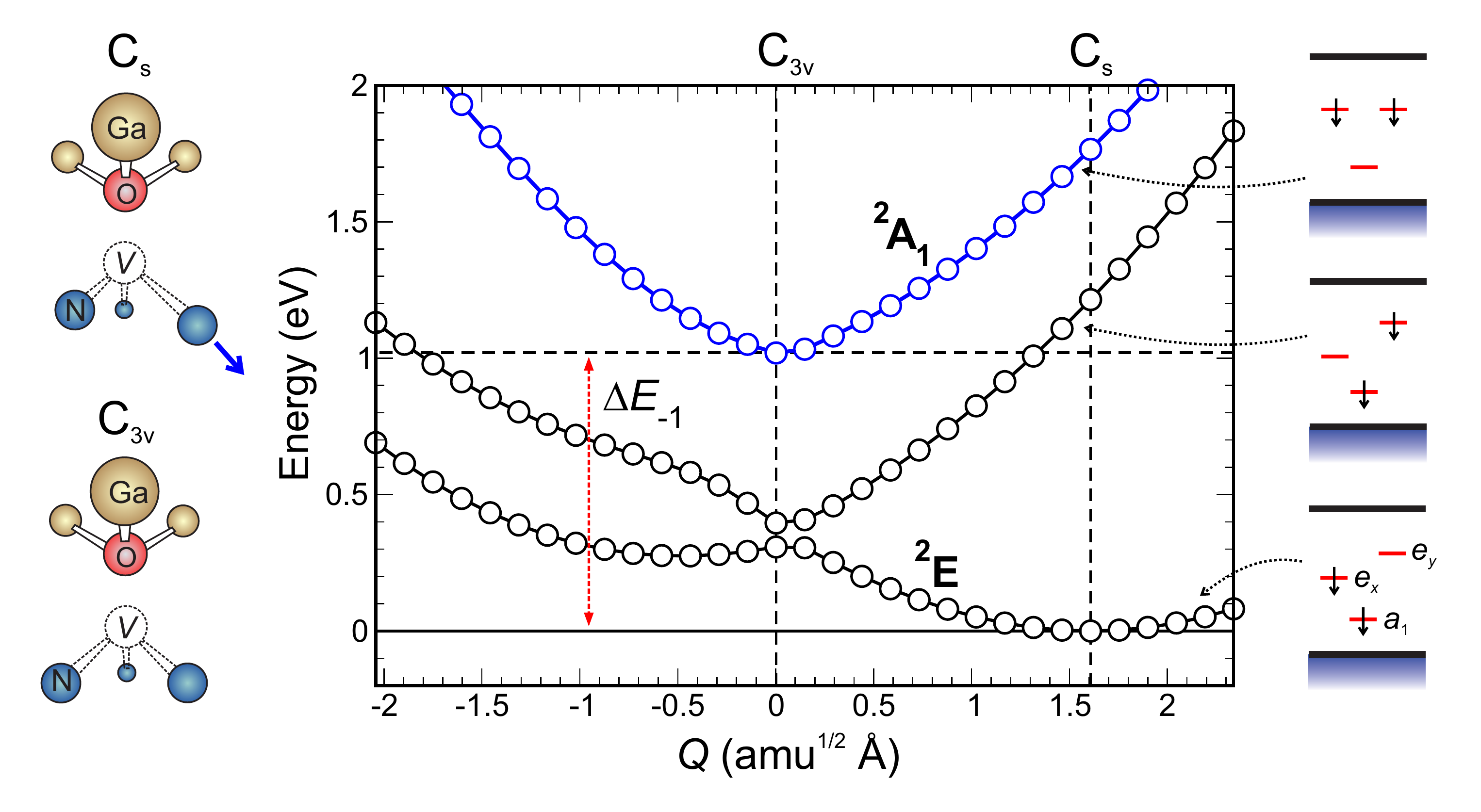}
\caption{1D configuration coordinate diagram of the VO defect in the negative charge state. The configuration coordinate interpolates between the equilibrium geometry of the ground state 
and a $C_{3v}$ geometry 
Ball-and-stick models on the left geometries of the defect in these two configurations. Solid arrows indicate relaxations of N atoms. Figures on the right show single-particle states and occupations in the minority spin channel for each of the electronic configurations. The $^2E$ electronic state with lower energy has the lower $e$ state filled, while the one with the higher energy has the higher $e$ state filled. They are degenerate for the $C_{3v}$ symmetry. In the $^2A_1$ electronic state the $a_1$ defect state is empty, while both $e$ states are filled.
}
\label{ccd-1}
\end{figure*}

To determine the excited-state energy of the $-1$ charge state we follow a similar strategy.
The ground state has an electronic configuration $a_1^1e^1$, and therefore it must also be a Jahn-Teller system with two branches. Were the symmetry $C_{3v}$, the total wavefunction would transform as $^2E$.
The first ``excited'' state (lowest-energy excitation) corresponds to removing an electron from a lower-lying $e$ state and putting it into a higher lying $e$ state (see Fig.~\ref{ccd-1}). Following the same reasoning as for the neutral state, we can show that its energy is given by:
\begin{equation}
E_2^{q=-1} = E_1^{q=-1} + \varepsilon_3^{q=0}-\varepsilon_2^{q=0}.
\end{equation}
However, this is not a real excited state, but simply the other branch of the $^2E$ manifold. 

The excited state we are looking for corresponds to removing an electron from the $a_1$ orbital (which is the second highest-occupied orbital) and putting it into a higher-lying empty $e$.  The electron removal energy is given by $\varepsilon_1^{q=-1}$. The electron is added to a system with a configuration $a_1^0e^1$, which is the excited state $0^*$ of the neutral defect discussed above, whereby defect states 1 and 3 are empty, and state 2 is filled. If we could calculate such a state, the electron addition energy would be $\varepsilon_3$ that corresponds to $0^*$. We approximate this value by $\varepsilon_3^{q=0}$, whereby state 1 is filled, while states 2 and 3 are empty. Therefore:
\begin{equation}
E_3^{q=-1} = E_1^{q=-1} + \varepsilon_3^{q=0}-\varepsilon_1^{q=-1}.
\end{equation}

Excitation energies were determined for configurations that linearly interpolate between two geometries. One geometry is the ground-state configuration of the $-1$ charge state, in which one N atom experiences a strong outward relaxation, as shown in Fig.~\ref{ccd-1}. This is again a $C_s$ geometry. Another geometry is a $C_{3v}$ geometry, which corresponds to an equilibrium configuration of the defect with all defect states occupied by $2/3$ of an electron (and is therefore a different geometry than in the case of the neutral center; the difference between the two is a breathing relaxation of the nitrogen neighbors). The results for energies are presented in Fig.~\ref{ccd-1}. The ground state indeed corresponds to a doubly-degenerate $^2E$ manifold, clearly showing the signatures of the $E \otimes e$ Jahn-Teller system \cite{Bersuker}: for the configuration coordinate shown in Fig.~\ref{ccd-1} the minimum occurs to the right of the symmetric configuration, while the transition state between the other two equivalent minima of the $E \otimes e$ Jahn-Teller system is seen as a local minimum occurring to the left of the symmetric configuration. The excited state is $^2A_1$ with an equilibrium geometry very close to a symmetric $C_{3v}$ configuration, as expected from the symmetry of the wavefunction.
The excitation energy is found to be $\Delta E_{-1}=1.02$ eV.

In both the neutral and the negative charge state the excited state is obtained by promoting the $a_1$ electron to the $e$ state, and this explains why $\Delta E_0$ is similar to $\Delta E_{-1}$. We observe that the $a_1$-$e$ splitting slightly decreases in expanded GaN (used to model InGaN, see main text and Sec.~\ref{InGaN}) with respect to unstrained GaN . We obtain energies of excited states in In$_{x}$Ga$_{1-x}$N by subtracting the decrease of the $a_1$-$e$ splitting from $\Delta E_0$ and $\Delta E_{-1}$. This results in very small relative changes of the excited-state energies. For example, for $x=0.3$ (band gap 2.34 eV), $\Delta E_0$ and $\Delta E_{-1}$ decrease by 0.12 eV.

\section{Intra-defect relaxation rate}
\label{sec:rel}

Like carrier capture, intra-defect transitions can occur radiatively or nonradiatively, and the two rates add up. For the case of carrier capture (transition from a band edge to a defect level), radiative processes can be neglected as a contribution to Shockley-Read-Hall recombination in group-III nitrides (see arguments in the main text). This has to be re-evaluated for intra-defect transitions.

The optical transitions $^3$$E \rightarrow $$^3$$A_2$ (for the neutral defect) and $^2$$A_1 \rightarrow $$^2$$E$ (for the negatively charged defect) are dipole-allowed. Within the Franck-Condon approximation the radiative transition rate is given by \cite{Ridley}
\begin{equation}
\Gamma= \frac{e^2 n_{r}}{3m^2\varepsilon_0\pi c^3 \hbar^2}g\left|\vec{p}_{if}\right|^2\Delta E_{opt},
\label{rate}
\end{equation}
where $n_\text{r}$ is the index of refraction ($n_\text{r}$=2.3 for GaN), $\Delta E_{opt}=m\omega$ is the energy of the transition, $g$ is the degeneracy of the final state ($2$ for the negatively charged defect, $1$ for the neutral one), and $\vec{p}_{if} =\Braket{\psi_{i}(x) |-i\hbar\vec{\nabla}|\psi_{f}(x)}$ is the momentum matrix element of the transition (related to the transition dipole moment $r_{if}$ via $p_{if}=m\omega r_{if}$). We obtain $\Gamma = 3.2\times10^{7}$ s$^{-1}$ for the neutral defect, and $\Gamma = 3.1\times10^{7}$ s$^{-1}$ for the negatively charged defect. These rates put a lower bound on the overall transition rate.

Evaluation of nonradiative rates is more complicated. Within the first order of electron-phonon coupling the nonradiative intra-defect relaxation rate is given by an expression analogous to Eq.~(1) of the main text \cite{DiBartolo}:
\begin{equation}
r=\frac{2\pi}{\hbar}gW_{if}^2\sum_{m,n} w_m (T)
\left| \left \langle\chi_{im}|Q+\Delta Q|\chi_{fn}\right \rangle \right |^2
\delta (\Delta E + m\hbar\Omega_{i}-n\hbar\Omega_{f}).
\label{eq:r}
\end{equation}
Here $W_{if}=6.4\times 10^{-2}$ eV/(amu$^{1/2}$\AA) is our calculated electron-phonon coupling matrix element between the $e$ and the $a_1$ state. Since Eq.~(\ref{eq:r}) is based on the harmonic approximation, we need to use a parabolic approximation for the potential energy surface, both in the ground state and in the excited state.  As is clear from, e.g.,  the configuration coordinate diagram in Fig.~\ref{ccd0}, the anharmonicity is significant. Using the parabolic approximation for the $^3A_2$ state in the region $Q<0$, and using the right branch of the $^3E$ excited state, we arrive at a rate larger than $10^{10}$ s$^{-1}$, which is significantly faster than the radiative relaxation.

Experimental information about spin-conserving intra-defect transitions in $sp$-bonded materials is scarce. However, a few recent well-studied examples confirm that nonradiative intra-defect relaxation at infrared energies is very fast \cite{Rogers_PP_2010}. For example, the ${^2}E_u \rightarrow {^2}E{_g}$ transition at the negatively charged silicon-vacancy center in diamond, with an energy separation of 1.68 eV, occurs at a rate $7.8\times10^{8}$ s$^{-1}$ at room temperature \cite{Rogers_Nature_2014}. The $^1E \rightarrow {^1}A_1$ transition in the negatively charged nitrogen-vacancy center in diamond, with an energy separation of 1.19 eV, is even faster and occurs at a rate larger than $10^9$ s$^{-1}$ \cite{Acosta_PRB_2010}. These examples support our observation that for wide-band-gap semiconductors spin-conserving nonradiative intra-defect transitions in the energy range around 1 eV are very fast ($10^{9}$ s$^{-1}$ or higher), justifying our assumption that intra-defect relaxations rates for the VO defect are high enough not to be rate-limiting.

\end{document}